\documentclass[prc,aps,eqsecnum,floatfix,showpacs]{revtex4-1}
\usepackage{color}
\usepackage{amssymb}
\usepackage{amsmath}
\usepackage{graphicx}
\usepackage{amsfonts}
\usepackage{slashed}
\usepackage{pstricks}
\usepackage{float}
\usepackage{hyperref}
\usepackage{array}
\begin{document}
\title{Scaling in the short-time approximation}
\author{
S. Pastore$^{ {\rm a} }$,
J. Carlson$^{ {\rm b} }$,
S. Gandolfi$^{ {\rm b} }$,
R. Schiavilla$^{ {\rm c,d} }$, and
R. B. Wiringa$^{ {\rm e}} $}

\affiliation{
$^{\,{\rm a}}$\mbox{Department of Physics and the McDonnell Center for the Space Sciences at Washington 
University in St. Louis, MO 63130}
$^{\,{\rm b}}$\mbox{Theoretical Division, Los Alamos National Laboratory, Los Alamos, NM 87545}\\
$^{\,{\rm c}}$\mbox{Theory Center, Jefferson Lab, Newport News, VA 23606}\\
$^{\,{\rm d}}$\mbox{Department of Physics, Old Dominion University, Norfolk, VA 23529}
$^{\,{\rm e}}$\mbox{Physics Division, Argonne National Laboratory, Argonne, IL 60439}\\
}
\date{\today}

\begin{abstract}
We briefly review the concept of
scaling and how it occurs in quasielastic electron and neutrino scattering from nuclei, and then the particular
approach to scaling in the short-time approximation.  We show that, while two-nucleon currents do
significantly enhance the transverse electromagnetic response, they do not spoil scaling, but in fact
enhance it.  We provide scaling results obtained in the short-time approximation that verify this claim.
The enhanced scaling is not ``accidental''---as claimed in Ref.~\cite{Benhar:2020jye}---but rather reflects
the dominant role played by pion exchange interactions and currents in the quasielastic regime.
\end{abstract}

\maketitle

\section{Quasielastic scattering and scaling in nuclear physics}

The concept of scaling originated from the idea to treat the
response functions as arising from the incoherent sum of scattering
from single nucleons.  Ignoring final-state interactions of the struck particle
with the rest of the nucleus yields final states that are products of free single-particle
states with the momenta of the original nucleon plus the transferred momentum,
and of the final states of the remaining interacting nucleons.  Ignoring
final-state-interaction effects in the final system altogether or including the
removal energy of the struck nucleon results in the
plane-wave impulse approximation (PWIA)~\cite{Benhar:2005dj,Benhar:2004mq} or the 
spectral function approach~\cite{Rocco:2015cil,Benhar:1994hw,Benhar:1998zd,Vagnoni:2017hll}.
The latter is an improvement as it contains additional information on the
energy to remove a nucleon from the nucleus.

While these approximations are useful to get an initial picture
of quasielastic scattering, they are incomplete.  They
predict, for example, that the longitudinal and transverse scaling functions
obtained from the corresponding response functions would be the same,
which is not observed experimentally.  For example, Fig.~30 of the
review article by Benhar, Day, and Sick~\cite{Benhar:2008}
shows that the scaling functions extracted from longitudinal and
transverse data on $^{12}$C differ in magnitude by approximately 40\%
for momentum transfers in the range $q\,$=$\,400$--600 MeV/c across the entire
quasielastic region.  Calculations in subsequent (as well as previous)
years~\cite{Carlson:1994,Carlson:2002,Pastore:2020,Lovato:2020}
have demonstrated that this dramatic difference arises largely due to the
interference between processes involving single-nucleon currents with an accompanying
correlated nucleon and processes involving two-nucleon currents.  Such
interference might be considered beyond the scope of the traditional scaling
approach, however we demonstrate that it can be accommodated within the
scaling picture.  We note that scaling of the individual
longitudinal and transverse response functions with energy and momenta provides
a much better description of the data than using a single scaling function
for both.

\section{The short-time Approximation and scaling}

As discussed in our paper~\cite{Pastore:2020}, quasielastic scaling is dominated by relatively
high momentum and energy scales, larger than the typical Fermi momentum and
Fermi energy.  In such a regime, the electroweak response is dominated
by nearly local quantities that can be calculated in terms of the one-
and two-body off diagonal density matrix and single- and two-nucleon currents.
This is natural in the path-integral picture, since high energies correspond
to short propagation times and hence small distances.

The short-time approximation is based on this path integral picture,
and scaling will naturally occur, both the $y$-scaling associated with the
response at different momentum transfer and the super-scaling associated
with the response of different nuclei. The latter is a consequence of the locality
of inclusive scattering.  The short-time approximation reduces to the plane-wave
impulse approximation if we ignore {\it i}) coherent scattering terms where two different
nucleons are struck, {\it ii}) two-nucleon currents, and {\it iii}) interactions in the two-nucleon
propagator.  In this simplification the calculation reduces to calculating
the one-body off diagonal density matrix or the momentum distribution.

While the standard PWIA or spectral function might superficially appear to be
independent of two-nucleon dynamics, they very much depend on the momentum
distribution in the ground state.  As demonstrated both theoretically~\cite{Schiavilla:2006xx,Cruz-Torres:2019fum}
and experimentally~\cite{Subedi:2008zz,Hen614}, the momentum distribution in the range of
$k \approx 2$ fm$^{-1}$ is dominated by the two-nucleon pion-exchange interaction.
So from this perspective it could be argued that PWIA does not just reflect single-nucleon
dynamics.

To go beyond this simplified picture, the STA  includes interactions in the two-nucleon
propagator.  The two-nucleon propagation is no longer a function only of the 
distance between the initial and propagated nucleon, it also depends strongly on
the spin and isospin of the nucleons.  Indeed, the response can be quite different 
depending on the nature of the single-nucleon coupling, and responses obtained with
a number of different such couplings are compared in \cite{Pandharipande:1994}.
For example, in the longitudinal channel charge can propagate through the exchange
of charged pions in addition to the momentum of the struck proton.  This additional
mechanism again defies the notion of longitudinal scattering as being driven only
by single-nucleon dynamics.  It turns out that such a mechanism gives a significant
contribution to the energy weighted sum rule, as the vertex and the Hamiltonian do
not commute, and produces a redistribution of strength, and in particular a larger tail in the
response at energies above the quasielastic peak.

\begin{figure}[h]
\centering
\includegraphics[width=0.7\columnwidth]{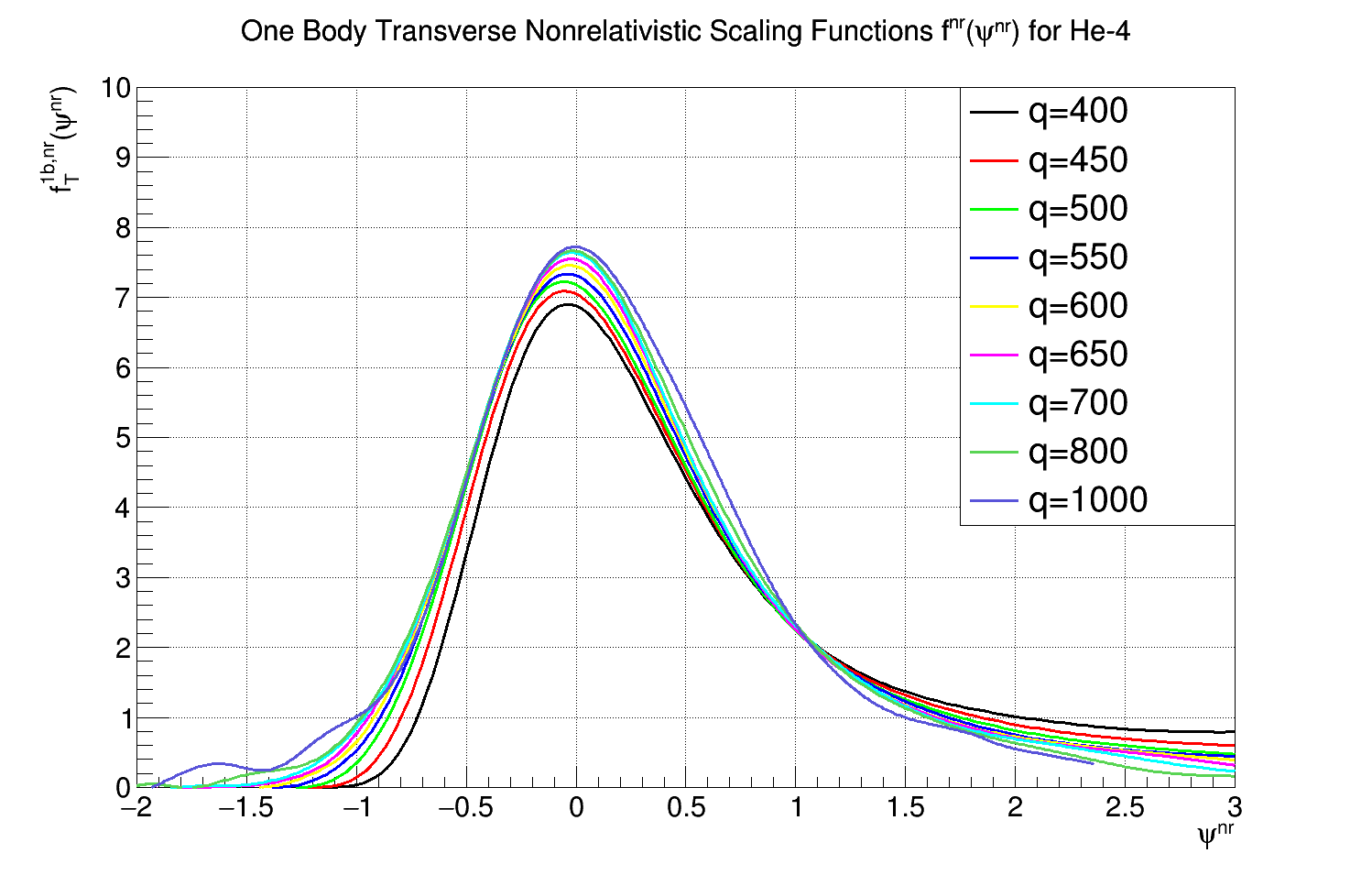} 
\caption{Transverse scaling functions for $^4$He with single-nucleon currents only.
Figure from Barrow {\it et al.}~\cite{Barrow}.}
\label{fig:scaleone}
\end{figure}

\begin{figure}[h]
\centering
\includegraphics[width=0.7\columnwidth]{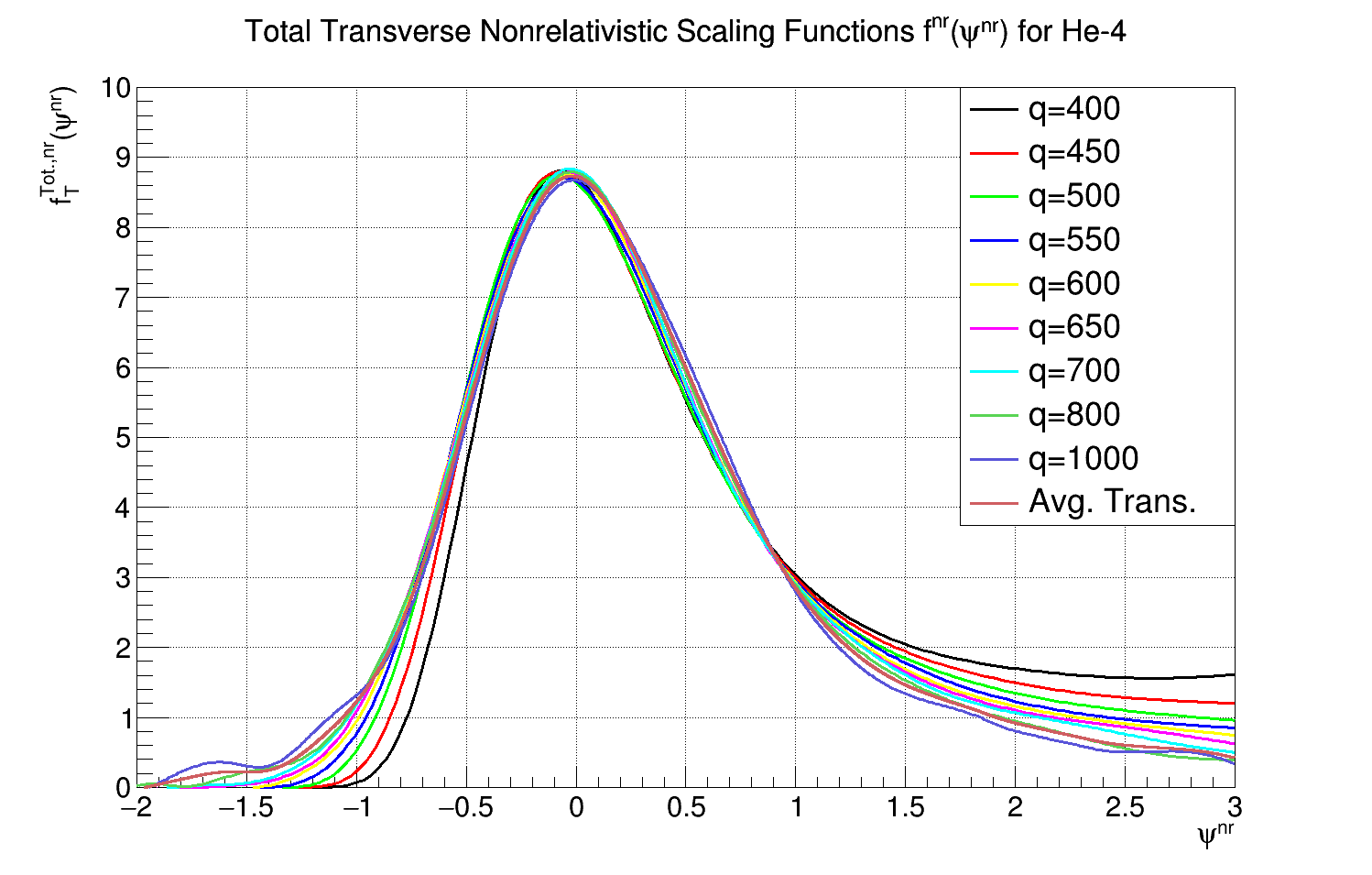} 
\caption{Transverse scaling functions for $^4$He with one- and  two-nucleon 
currents. Figure from Barrow {\it et al.}~\cite{Barrow}.}
\label{fig:scaletwo}
\end{figure}

We demonstrate the scaling properties of the STA response functions in Figs.~\ref{fig:scaleone}
and~\ref{fig:scaletwo} for the transverse channel in ${^4}$He.
Figure~\ref{fig:scaleone} shows the scaling function at various momentum
transfers using only single-nucleon current operators.  The scaling is
reasonable but not perfect as additional mechanisms
(incoherent scattering, final state interactions, etc.) come into play.

The next figure (Fig.~\ref{fig:scaletwo}) shows the scaling function
including two-nucleon currents.  The most important two-nucleon currents
in this regime are those due to pion exchange.  The scaling function is enhanced
compared to that obtained with single-nucleon currents only due to the
interference described in our paper.  More importantly, the scaling with
momentum transfer is in fact substantially better with the inclusion of
two-nucleon currents.

As we describe in our paper~\cite{Pastore:2020}, the momentum structure of the pion-exchange
in the strong interaction arises in exactly the same way as the momentum structure
in the pion-exchange two-nucleon currents.  Hence the scaling is no less
apparent in the full calculation as in the calculation with single-nucleon
currents only.   In our view, calling this scaling ``accidental''~\cite{Benhar:2020jye} is akin
to calling the relation between two-nucleon currents and two-nucleon interactions
``accidental'' when in fact it is governed by the same underlying dynamics of pion
exchange mechanisms, an essential feature in describing quasielastic scattering.

\vspace{0.5cm}

The present research is supported by the U.S. Department of Energy, Office of Science,
Office of Nuclear Physics, under contracts DE-SC0021027 (S.P.),  DE-AC02-06CH11357 and DE-AC52-06NA25396
(S.G. and J.C.), DE-AC05-06OR23177 (R.S.), and DE-AC02-06CH11357 (R.B.W.), and the
U.S. Department of Energy funds through the FRIB Theory Alliance award DE- SC0013617 
and through the Neutrino Theory Network (NTN)
(S.P.).

\bibliography{biblio}
\end{document}